\documentclass[preprint,eqsecnum,aps,nofootinbib]{revtex4}
\usepackage{amsfonts,amsmath,amssymb,amsthm}
\usepackage{latexsym}
\usepackage{bbm,bm}
\usepackage{graphicx}

\newcommand{\beq}{\begin{equation}}
\newcommand{\eeq}{\end{equation}}
\newcommand{\beqs}{\begin{eqnarray}}
\newcommand{\eeqs}{\end{eqnarray}}

\begin{document}

\title{ Aharonov-Bohm-Like Scattering in the Generalized Uncertainty Principle-corrected Quantum Mechanics}

\author{DaeKil Park$^{1,2}$\footnote{dkpark@kyungnam.ac.kr} }

\affiliation{$^1$Department of Electronic Engineering, Kyungnam University, Changwon
                 631-701, Korea    \\
             $^2$Department of Physics, Kyungnam University, Changwon
                  631-701, Korea    
                      }

\begin{abstract}
We discuss classical electrodynamics and the Aharonov-Bohm effect in the presence of the minimal length. In the former we derive the classical equation of motion and the 
corresponding Lagrangian. In the latter we adopt the generalized uncertainty principle (GUP) and compute the scattering cross section up to the first-order of the 
GUP parameter $\beta$. Even though the minimal length exists, the cross section is invariant under the simultaneous change $\phi \rightarrow -\phi$, $\alpha' \rightarrow -\alpha'$, where 
$\phi$ and $\alpha'$ are azimuthal angle and magnetic flux parameter. However, unlike the usual Aharonv-Bohm scattering the cross section exhibits discontinuous behavior at every 
integer $\alpha'$. The symmetries, which the cross section has in the absence of GUP, are shown to be explicitly broken at the level of ${\cal O} (\beta)$.  
\end{abstract}

\maketitle

\section{Introduction}

The most theories of quantum gravity predict the existence of a minimal length \cite{mead64,townsend76,amati89,garay94} at the Planck scale. It appears as various different expressions in loop quantum gravity\cite{rovelli98,carlip01},
string theory\cite{konishi90,kato90}, path-integral quantum gravity\cite{padmanabhan85,padmanabhan87,greensite91}, and black hole physics\cite{maggiore93}. From the aspect of quantum mechanics the existence of a minimal 
length results in the modification of the Heisenberg uncertainty principle (HUP)\cite{uncertainty,robertson1929} $\Delta P \Delta Q \geq \frac{\hbar}{2}$, because $\Delta Q$ should be larger than the minimal length.
Various modification of HUP, called the generalized uncertainty principle (GUP), were suggested in Ref. \cite{kempf93,kempf94}.
The GUP has been used to explore the various branches of physics such as micro-black hole\cite{scar99-1}, gravity\cite{adler99-1}, 
cosmological constant\cite{okamura02-1}, and classical central potential problem\cite{okamura02-2}. It is also used in the 
low-energy regime\cite{scar10-1} and the emergence of (doubly) special relativity\cite{scar12-1}. The experimental detection of GUP  was emphasized 
in Ref. \cite{scar15-1}, where GUP is directly linked to the deformation of the spacetime metric. In this way the existence of GUP 
can be experimentally verified by measuring the light deflection and perihelion procession. As we will show in this paper, the 
existence of GUP also can be verified by measuring the cross section of the Aharonov-Bohm-Like scattering. 

In this paper\footnote{Although the effect of the presence of a minimal length should be discussed in a relativistic fashion, we will examine it in the non-relativistic quantum mechanics when the Aharonov-Bohm potential is 
involved. Therefore, the results presented in this paper should be modified when the relativistic effect is included.}  we will choose the $d$-dimensional GUP in a form
\begin{equation}
\label{GUP-d-1}
\Delta P_i \Delta X_i \geq \frac{\hbar}{2} \left[ 1 + \beta \left(\Delta {\bf P}^2 + \langle {\bf P} \rangle^2 \right) + 2 \beta \left( \Delta P_i^2 + \langle P_i \rangle^2 \right) \right]
\hspace{1.0cm}  (i = 1, 2, \cdots, d)
\end{equation}
where $\beta$ is a GUP parameter, which has a dimension $(\mbox{momentum})^{-2}$. Using $\Delta A \Delta B \geq \frac{1}{2} | \langle [A, B] \rangle |$, Eq. (\ref{GUP-d-1}) induces the 
modification of the commutation relations as\footnote{One may wonder whether the commutation relations (\ref{GUP-d-2}) are inconsistent with 
each other. If, in fact, we impose $[P_i, P_j] = 0$, the Jacobi identity determines $[X_i, X_j]$ in a form 
$$[X_i, X_j] = \frac{4 \beta^2 {\bf P}^2}{1 + \beta {\bf P}^2} (P_i X_j - P_j X_i) = {\cal O} (\beta^2).$$
Since we will explore the AB-like phenomenon up to first of $\beta$, Eq. (\ref{GUP-d-2}) is valid for this reason.} 
\begin{eqnarray}
\label{GUP-d-2}
&& \left[ X_i, P_j \right] = i \hbar \left( \delta_{ij} + \beta \delta_{ij} {\bf P}^2 + 2 \beta P_i P_j  \right)    \\    \nonumber
&& \hspace{1.0cm} \left[ X_i, X_j \right] = \left[P_i, P_j \right] = 0.
\end{eqnarray}

\begin{figure}[ht!]
\begin{center}
\includegraphics[height=6.0cm]{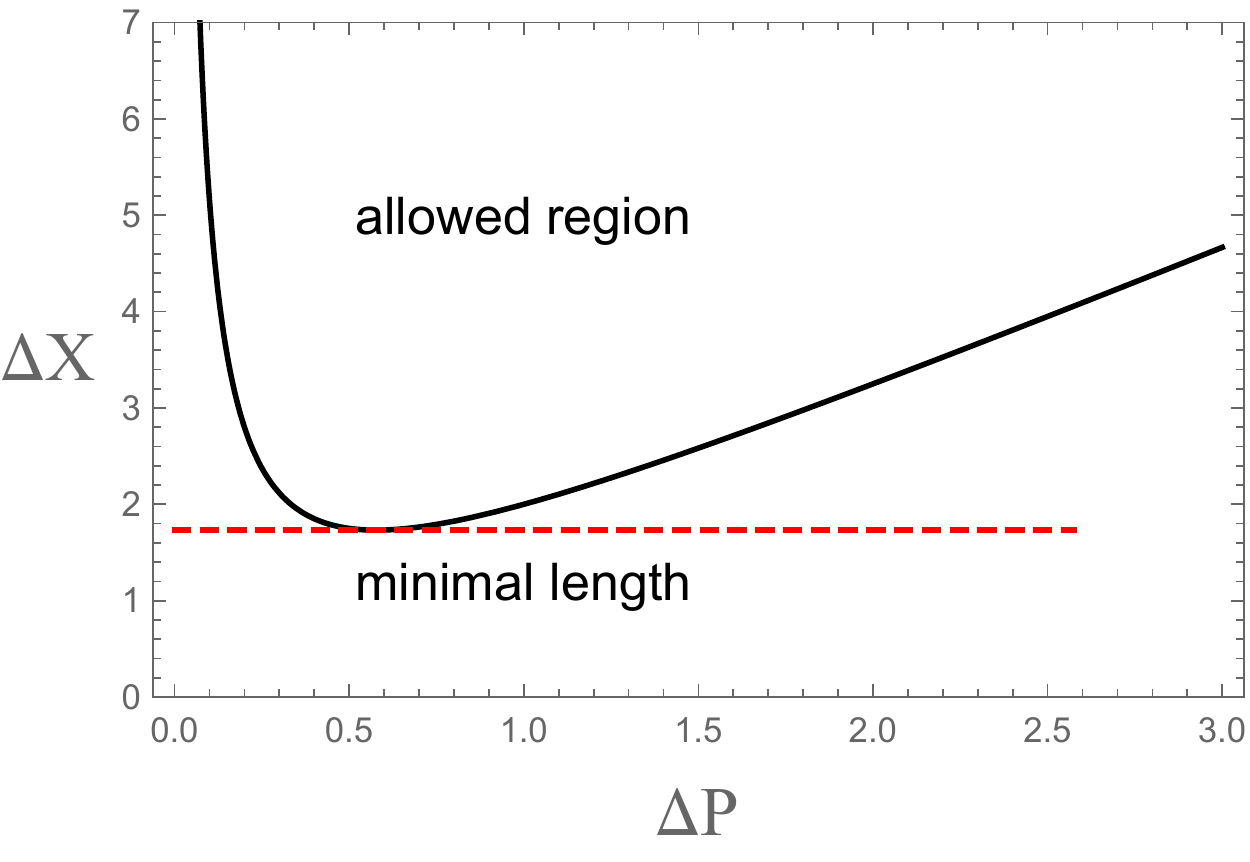} 

\caption[fig1]{(Color online) The minimal length and allowed region of one-dimensional GUP (\ref{GUP-d-3}) when $\hbar = \beta = 1$. 
 }
\end{center}
\end{figure}

The existence of the minimal length is easily shown at $d = 1$. In this case  Eq. (\ref{GUP-d-1}) is expressed as 
\begin{equation}
\label{GUP-d-3}
\Delta P \Delta X \geq \frac{\hbar}{2} \left( 1 + 3 \beta \Delta P^2 \right)
\end{equation}
if $\langle P \rangle = 0$. Then, the equality of Eq. (\ref{GUP-d-3}) yields 
\begin{equation}
\label{minimal-length}
\Delta X^2 \geq \Delta X_{min}^2 = 3 \beta \hbar^2.
\end{equation}
In Fig. 1 the allowed region and minimal length of Eq. (\ref{GUP-d-3}) is plotted when $\hbar = \beta = 1$. 

If $\beta$ is small, Eq. (\ref{GUP-d-2}) can be solved as 
\begin{equation}
\label{GUP-d-4}
P_i = p_i \left(1 + \beta {\bf p}^2 \right) + {\cal O} (\beta^2)   \hspace{1.0cm} X_i = x_i
\end{equation}
where $p_i$ and $x_i$ obey the usual HUP. Using Eq. (\ref{GUP-d-4}) and Feynman's path-integral technique\cite{feynman,kleinert} the Feynman propagator (or kernel) was exactly derived 
up to ${\cal O} (\beta)$ for $d=1$ free particle case\cite{das2012,gangop2019}. Also the propagator for  $d$-dimensional simple harmonic oscillator system was also derived recently in Ref. \cite{comment-1,park20-1}.

The main purpose of this paper is to examine how the Aharonov-Bohm (AB) effect\cite{AB-1,hagen-91} is modified when the GUP (\ref{GUP-d-1}) is introduced. 
The AB effect is a pure quantum mechanical phenomenon, which predicts that the electromagnetic vector potential plays a role of observable at the quantum level when the charged particle moves around an infinitely thin magnetic flux tube. 
The experimental realization of this effect was discussed in Ref. \cite{peshkin}.
The effect of the particle spin in the 
AB-scattering was examined a few years ago in Ref. \cite{hagen-91,hagen-90-2,park-95}. In particular, when the spin is $1/2$, the corresponding  Schr\"{o}dinger-like equation derived from Dirac equation  
involves the $\delta$-function potential\cite{hagen-91,jackiw} as a Zeeman interaction.  In order to make the theory finite a mathematically-oriented self-adjoint extension\cite{capri} or the physically-oriented renormalization\cite{huang} can be adopted. 
The equivalence of both methods was discussed in Ref. \cite{jackiw,park97-1}. 

The paper is organized as follows. In the next section we derive the classical equation of motion up to ${\cal O} (\beta)$ by making use of the Poisson bracket formalism when the minimal length (\ref{minimal-length}) exists.
Also, the classical  Lagrangian is explicitly derived in this section. In section III we discuss the AB-scattering in the presence of GUP (\ref{GUP-d-4}). Unlike the usual AB-effect with HUP it is shown that the irregularity at the origin
cannot be avoided because of the effect of GUP. The scattering cross section is shown to be discontinuous at every integer $\alpha' = \alpha / \hbar$, where $\alpha$ is a magnetic flux parameter. The various symmetries of the cross section 
in the usual AB-scattering are explicitly broken. In section IV a brief conclusion is given.

\section{Classical Electrodynamics in the presence of the minimal length}

In this section we discuss how the classical electrodynamics is modified if the minimal length (\ref{minimal-length}) exists. We start with a classical Hamiltonian
\begin{equation}
\label{class-hamil-1}
H_{cl} = \frac{1}{2 M} \left( \bm{P} - q \bm{A} \right)^2 + q V = H_{0,cl} + \frac{\beta}{M} \bm{p}^2 \left( \bm{p} - q \bm{A} \right) \cdot \bm{p} + {\cal O} (\beta^2)
\end{equation}
where $\bm{A}$ and $V$ are the vector and scalar potentials, and $H_{0,cl}$ is the classical Hamiltonian when there is no minimal length, which is explicitly expressed by 
\begin{equation}
\label{class-hamil-2}
H_{0,cl} = \frac{1}{2 M} \left( \bm{p} - q \bm{A} \right)^2 + q V.
\end{equation}
In Eq. (\ref{class-hamil-1}) we used Eq. (\ref{GUP-d-4}). Of course, we have not considered the ordering problem of $\bm{p}$ and $\bm{x}$ because we deal with the classical Hamiltonian. 

In order to derive a classical equation of motion we use the Poisson bracket
\begin{equation}
\label{poisson-1}
\dot{x}_j \equiv \{H_{cl}, x_j \} = \frac{\partial H_{cl}}{\partial p_j} \frac{\partial x_j}{\partial x_j} - \frac{\partial H_{cl}}{\partial x_j}  \frac{\partial x_j}{\partial p_j},
\end{equation}
which yields
\begin{equation}
\label{poisson-2}
M \dot{\bm{x}} = \bm{p} - q \bm{A} + \beta \left[4 \bm{p}^2 \bm{p} - 2 q (\bm{A} \cdot \bm{p}) \bm{p} - q \bm{p}^2 \bm{A} \right] + {\cal O} (\beta^2).
\end{equation}
Also, one can compute $\dot{p}_j = \left\{H_{cl}, p_j \right\}$, which gives
\begin{equation}
\label{poisson-3}
\dot{p}_j = \left[ \frac{q}{M} (\bm{p} - q \bm{A}) \cdot \frac{\partial \bm{A}}{\partial x_j} - q \frac{\partial V}{\partial x_j} \right] 
+ \frac{\beta q}{M} \bm{p}^2 \left( \frac{\partial \bm{A}}{\partial x_j} \cdot \bm{p} \right) + {\cal O} (\beta^2).
\end{equation}

Combining Eqs. (\ref{poisson-2}) and (\ref{poisson-3}) with long and tedious calculation, it is possible to derive
\begin{equation}
\label{classical-1}
M \ddot{x}_j = q (\bm{E} + \bm{v} \times \bm{B} )_j + \beta q \Gamma_j + {\cal O} (\beta^2)
\end{equation}
where $\bm{E} = - \bm{\nabla} V - \frac{\partial}{\partial t} \bm{A}$ and $\bm{B} = \bm{\nabla} \times \bm{A}$, which are the usual electric and magnetic fields.
The correction term $\Gamma_j$ at the first order of $\beta$ is expressed as 
\begin{eqnarray}
\label{classical-2}
&&\Gamma_j = \bm{H}_{11}^2 (\bm{E} + \bm{v} \times \bm{B} )_j + 2 H_{11,j} (\bm{H}_{11} \cdot \bm{E}) -2 m v_j (\bm{H}_{32} \cdot \bm{\nabla} V) - 2 q A_j (\bm{H}_{21} \cdot \bm{\nabla} V)       \nonumber   \\
&& \hspace{2.0cm} - \frac{\partial V}{\partial x_j} (\bm{H}_{11} \cdot \bm{H}_{31}) + 2 q H_{32,j} \left(\bm{A} \cdot (\bm{v} \times \bm{B}) \right) + 2m H_{32,j} (\bm{G} \cdot \bm{v})        \\    \nonumber
&&  \hspace{3.0cm}+ 2 q H_{21,j} (\bm{G} \cdot \bm{A}) - q (\bm{v} \cdot \bm{H}_{11}) \frac{\partial}{\partial x_j} \bm{A}^2
\end{eqnarray}
where
\begin{equation}
\label{classical-3} 
\bm{H}_{a b} = a m \bm{v} + b q \bm{A}   \hspace{1.0cm} \bm{G} = \sum_{i} v_i \frac{\partial \bm{A}}{\partial x_i}.
\end{equation}
The equation of motion (\ref{classical-1}) can be derived as an Euler-Lagrange equation  from the Lagrangian
\begin{equation}
\label{classical-4}
L (\bm{x}, \dot{\bm{x}}) = \frac{1}{2} m \dot{\bm{x}}^2 + q \dot{\bm{x}} \cdot \bm{A} - q V - \beta \bm{H}_{11}^2 (\dot{\bm{x}} \cdot \bm{H}_{11}) + {\cal O} (\beta^2),
\end{equation}
where $\bm{v}$ in $\bm{H}_{a b}$ should be replaced by $\dot{\bm{x}}$ in Eq. (\ref{classical-4}). Unlike the classical equation of motion in the absence of the minimal length, the scalar and vector
potentials explicitly appear in Eq. (\ref{classical-1}) at the first order of $\beta$. This means that if the minimal length exists, the potentials $V$ and $\bm{A}$ are not merely mathematical tools for the 
derivation of $\bm{E}$ and $\bm{B}$ even at the classical level. Of course, the classical equation of motion (\ref{classical-1}) indicates that the usual gauge symmetry ${\bf A} \rightarrow {\bf A} + \frac{1}{q} \nabla \Lambda$ and 
$V \rightarrow V - \frac{1}{q} \frac{\partial \Lambda}{\partial t}$ does not hold at the classical level. However, one can show that this theory has a modified symmetry up to ${\cal O} (\beta)$ in a form:
\begin{equation}
\label{editor-1}
{\bf A} \rightarrow {\bf A} + \frac{1}{q} \left(\nabla \Lambda + \beta {\bf F} \right)  \hspace{1.0cm} V \rightarrow V - \frac{1}{q} \left( \frac{\partial \Lambda}{\partial t} + \beta F_0 \right)
\end{equation}
where ${\bf F}$ and $F_0$ satisfy 
\begin{equation}
\label{editor-2}
{\bf F} - 2 ({\bf H}_{11} \cdot \nabla \Lambda) {\bf H}_{11} - {\bf H}_{11}^2 \nabla \Lambda \equiv \nabla \Lambda_1    \hspace{1.0cm} F_0 = \frac{\partial \Lambda_1}{\partial t}.
\end{equation}
Under the transformation the Lagrangian (\ref{classical-4}) transforms $L \rightarrow L + \frac{d}{d t} \left( \Lambda + \beta \Lambda_1 \right)$. Of course, the symmetry (\ref{editor-1}) reduces to the usual gauge symmetry at $\beta = 0$. 
However, this symmetry is completely different from the usual one because the vector ${\bf H}_{11}$ contains not only particle's velocity but also the vector potential itself. 
The Lagrangian (\ref{classical-4}) can be used to explore the quantum electrodynamics in the presence of the minimal length
by applying the path-integral technique\cite{feynman,kleinert}.


\section{AB-like Phenomena with GUP}

In this section we examine how the AB effect\cite{AB-1} is modified when the GUP (\ref{GUP-d-4}) is introduced. 
The Hamiltonian with AB system can be written as 
\begin{equation}
\label{hamil-1}
\hat{H} = \frac{1}{2 M}\left(\bm{P} - e \bm{A} \right)^2 
= \frac{1}{2 M} \left[ (1 + \beta \bm{p}^2) \bm{p} - e \bm{A} \right]^2 + {\cal O} (\beta^2) = \hat{H}_0 + \beta \hat{H}_1 + {\cal O} (\beta^2),
\end{equation}
where $e$ is an particle charge and 
\begin{equation}
\label{additional-11}
\hat{H}_0 = \frac{1}{2 M} (\bm{p} - 2 \bm{A})^2   \hspace{.5cm}  \hat{H}_1 = \frac{1}{M} \left[(\bm{p}^2)^2 - \frac{e}{2} \left\{ (\bm{A} \cdot \bm{p}) \bm{p}^2 + \bm{p}^2 (\bm{p} \cdot \bm{A}) \right\} \right].
\end{equation}
If we represent the energy eigenvalue in terms of the wave number as $E = E_0 + \beta E_1 + {\cal O} (\beta^2)$ with $E_0 = \hbar^2 k^2 / (2 M)$ and $E_1 = \hbar^4 k^4 / M$,  
it is straightforward to show that the Schr\"{o}dinger equation $\hat{H} \psi = E \psi$ can be written as 
\begin{eqnarray}
\label{schrodinger-1}
&&(-i \hbar \bm{\nabla} - e \bm{A})^2 \psi + 2 \beta \hbar^4 \left[ (\bm{\nabla}^2)^2 - \frac{i e}{2 \hbar} \left( \bm{A} \cdot \bm{\nabla} \bm{\nabla}^2 + \bm{\nabla}^2 \bm{\nabla} \cdot \bm{A} \right) \right] \psi
                                                                                                                                                                                 \\    \nonumber
&&   \hspace{8.0cm}  + {\cal O} (\beta^2) = (\hbar^2 k^2 + 2 \beta \hbar^4 k^4)  \psi.
\end{eqnarray}

We assume that there is a thin magnetic flux tube along the $z$-axis, which gives the vector potential in a form:
\begin{equation}
\label{flux-1}
e A_i = \frac{\alpha \epsilon_{ij} x_j}{r^2},
\end{equation}
where $\epsilon_{ij}$ is an antisymmetric tensor with $\epsilon_{01} = - \epsilon_{10} = 1$. In the usual electromagnetic theory the choice of the vector potential (\ref{flux-1}) is not unique due to the gauge symmetry. 
Thus, the choice of Eq. (\ref{flux-1}) corresponds to the Coulomb gauge $\nabla \cdot {\bf A} = 0$. However, the gauge symmetry is modified to Eq. (\ref{editor-1}) for our case, which contains the particle's velocity. Thus, our 
results presented in the paper are valid only for the particular choice of ${\bf A}$ given in Eq. (\ref{flux-1}). 
Then, the corresponding magnetic field is 
$e B = - (\alpha / r) \delta (r)$. Using Eq. (\ref{flux-1}) explicitly, one can show 
\begin{equation}
\label{identity-1}
\bm{\nabla}^2 \bm{\nabla} \cdot (e \bm{A}) = (e \bm{A}) \cdot \bm{\nabla} \bm{\nabla}^2 + \frac{4 \alpha}{r^3} \left( \frac{\partial}{\partial r} - \frac{1}{r} \right) \frac{\partial}{\partial \phi}.
\end{equation}
Then, the Schr\"{o}dinger equation (\ref{schrodinger-1}) reduces to 
\begin{eqnarray}
\label{schrodinger-2}
&&\hspace{2.0cm} \left[ \frac{\partial^2}{\partial r^2} + \frac{1}{r} \frac{\partial}{\partial r} + \frac{1}{r^2} \left( \frac{\partial}{\partial \phi} + i \alpha' \right)^2 + k^2 \right] \psi          \\    \nonumber
&&- 2 \beta \hbar^2 \left[  (\bm{\nabla}^2)^2 + \frac{i \alpha'}{r^2} \left( \frac{\partial^2}{\partial r^2} - \frac{1}{r} \frac{\partial}{\partial r} + \frac{1}{r^2} \frac{\partial^2}{\partial \phi^2} 
                        + \frac{2}{r^2} \right) \frac{\partial}{\partial \phi} - k^4  \right] \psi + {\cal O} (\beta^2) = 0
\end{eqnarray}
where $\alpha' = \alpha / \hbar$. One can show that the Schr\"{o}dinger equation (\ref{schrodinger-2}) is invariant under the simultaneous operations $\alpha \rightarrow - \alpha$, $\phi \rightarrow - \phi$.

If one imposes
\begin{equation}
\label{impose-1}
\psi (\bm {r}) = \sum_{m=-\infty}^{\infty} e^{i m \phi} f_m (r),
\end{equation} 
the radial equation of (\ref{schrodinger-2}) can be written as 
\begin{equation}
\label{radial-1}
\left( \widehat{S}_1 - 2 \beta \hbar^2 \widehat{S}_2 \right) f_m (r) + {\cal O} (\beta^2) = 0,
\end{equation}
where
\begin{eqnarray}
\label{radial-2}
&&\widehat{S}_1 = \frac{d^2}{dr^2} + \frac{1}{r} \frac{d}{d r} - \frac{(m + \alpha')^2}{r^2} + k^2                             \\    \nonumber
&&\widehat{S}_2 = \left( \frac{d^2}{d r^2} + \frac{1}{r} \frac{d}{d r} - \frac{m^2 - \alpha' (m + \alpha')}{r^2} - k^2 \right) \widehat{S}_1              \\    \nonumber
&&\hspace{2.0cm}- \frac{2 \alpha' (3 m + 2 \alpha')}{r^2} \left( \frac{1}{r} \frac{d}{d r} - \frac{1}{r^2} + \frac{k^2}{2} \right)
+ \frac{\alpha'^2 (m + \alpha') (2 m + \alpha')}{r^4}.
\end{eqnarray}
The symmetry of the simultaneous operations $\alpha \rightarrow - \alpha$, $\phi \rightarrow - \phi$ is represented in the radial equation as the simultaneous changes $\alpha \rightarrow - \alpha$, $m \rightarrow - m$.
If we set 
\begin{equation}
\label{radial-3}
f_m (r) = f_{0,m} (r) + \beta f_{1,m} (r) + {\cal O} (\beta^2),
\end{equation}
within ${\cal O} (\beta)$ the radial equation is represented as the following two equations:
\begin{eqnarray}
\label{radial-4}
&&\widehat{S}_1 f_{0,m} (r) = 0                                                \\      \nonumber
&&\widehat{S}_1 f_{1,m} (r) = 2 \hbar^2 \left[ - \frac{2 \alpha' (3 m + 2 \alpha')}{r^2} \left( \frac{1}{r} \frac{d}{d r} - \frac{1}{r^2} + \frac{k^2}{2} \right)
+ \frac{\alpha'^2 (m + \alpha') (2 m + \alpha')}{r^4} \right] f_{0,m} (r).
\end{eqnarray}
The general solutions of Eq. (\ref{radial-4}) are 
\begin{eqnarray}
\label{radial-5}
&&f_{0,m} (r) = A_m J_{|m + \alpha'|} (z) + B_m J_{-|m + \alpha'|} (z)                  \\      \nonumber
&&f_{1,m} (r) = C_m J_{|m + \alpha'|} (z) + D_m J_{-|m + \alpha'|} (z) + u_m (z) J_{|m + \alpha'|} (z) + v_m (z) J_{-|m + \alpha'|} (z)
\end{eqnarray}
where $z = k r$ and $J_{\nu} (z)$ is usual Bessel function of the first kind. In Eq. (\ref{radial-5}) $u_m$ and $v_m$ are 
\begin{eqnarray}
\label{radial-6}
&&u_m (z) = \frac{\pi \hbar^2 k^2}{\sin (|m + \alpha'| \pi)}                                                                                         \\      \nonumber
&& \times \Bigg[ A_m \bigg\{ 2 \xi_m F_2 (z;-|m + \alpha'|, |m + \alpha'| + 1) + \xi_{m,-} F_3 (z;-|m + \alpha'|, |m + \alpha'|)                     \\     \nonumber
&& \hspace{8.0cm}    - \xi_m F_1 (z;-|m + \alpha'|, |m + \alpha'|) \bigg\}                                                                           \\     \nonumber
&& \hspace{.8cm} + B_m \bigg\{ 2 \xi_m F_2 (z;-|m + \alpha'|, -|m + \alpha'| + 1) + \xi_{m,+} F_3 (z;-|m + \alpha'|, -|m + \alpha'|)                   \\     \nonumber
&& \hspace{8.0cm}    - \xi_m F_1 (z;-|m + \alpha'|, -|m + \alpha'|) \bigg\}       \Bigg]                                                               \\     \nonumber
&&v_m (z) = -\frac{\pi \hbar^2 k^2}{\sin (|m + \alpha'| \pi)}                                                                                         \\      \nonumber
&& \times \Bigg[ A_m \bigg\{ 2 \xi_m F_2 (z;|m + \alpha'|, |m + \alpha'| + 1) + \xi_{m,-} F_3 (z;|m + \alpha'|, |m + \alpha'|)                     \\     \nonumber
&& \hspace{8.0cm}    - \xi_m F_1 (z;|m + \alpha'|, |m + \alpha'|) \bigg\}                                                                           \\     \nonumber
&& \hspace{.8cm} + B_m \bigg\{ 2 \xi_m F_2 (z;|m + \alpha'|, -|m + \alpha'| + 1) + \xi_{m,+} F_3 (z;|m + \alpha'|, -|m + \alpha'|)                   \\     \nonumber
&& \hspace{8.0cm}    - \xi_m F_1 (z;|m + \alpha'|, -|m + \alpha'|) \bigg\}       \Bigg] 
\end{eqnarray}
where 
\begin{equation}
\label{radial-7}
\xi_m = \alpha' (3 m + 2 \alpha')   \hspace{1.0cm} \xi_{m,\pm} = \alpha'^2 (m + \alpha') (2 m + \alpha') + 2 \xi_m (1 \pm |m + \alpha'|)
\end{equation}
and 
\begin{equation}
\label{radial-8}
F_n (z;\mu, \nu) \equiv \int \frac{J_{\mu} (z) J_{\nu} (z)}{z^n} d z.
\end{equation}
In order to escape the infinity at $r = 0$, we should choose $B_m = 0$. 
Therefore, the wave function can be written in a form;
\begin{eqnarray}
\label{wavef-1}
&&\psi ({\bm r}) = \sum_{m = -\infty}^{\infty} e^{i m \phi} A_m J_{|m + \alpha'|} (z) + \beta \sum_{m = -\infty}^{\infty} e^{i m \phi} \bigg( C_m J_{|m + \alpha'|} (z)                 \\      \nonumber 
&&  \hspace{1.0cm} + D_m J_{-|m + \alpha'|} (z)  + u_m (z) J_{|m + \alpha'|} (z) + v_m (z) J_{-|m + \alpha'|} (z)  \bigg) + {\cal O} (\beta^2)
\end{eqnarray}
where 
\begin{eqnarray}
\label{wave-2}
&&u_m (z) = \frac{A_m \pi \hbar^2 k^2 }{\sin (|m + \alpha'| \pi)} \bigg\{ 2 \xi_m F_2 (z;-|m + \alpha'|, |m + \alpha'| + 1)                           \\     \nonumber
&&\hspace{1.0cm}  + \xi_{m,-} F_3 (z;-|m + \alpha'|, |m + \alpha'|) - \xi_m F_1 (z;-|m + \alpha'|, |m + \alpha'|) \bigg\}                                                                \\     \nonumber
&&v_m (z) = -\frac{A_m \pi \hbar^2 k^2}{\sin (|m + \alpha'| \pi)} \bigg\{ 2 \xi_m F_2 (z;|m + \alpha'|, |m + \alpha'| + 1)                           \\     \nonumber
&&\hspace{1.0cm}  + \xi_{m,-} F_3 (z;|m + \alpha'|, |m + \alpha'|)  - \xi_m F_1 (z;|m + \alpha'|, |m + \alpha'|) \bigg\}.                                                                          
\end{eqnarray}

If we choose 
\begin{equation}
\label{coeff-1}
A_m = (-i)^{|m + \alpha'|} = e^{-i \pi |m + \alpha'| / 2},
\end{equation}
one can show\cite{hagen-91}
\begin{equation}
\label{scattering-1}
\lim_{r \rightarrow \infty} \sum_{m = -\infty}^{\infty} e^{i m \phi} A_m J_{|m + \alpha'|} (z) = e^{-i z \cos \phi} + \frac{e^{i k r}}{\sqrt{r}} f_0 (\phi)
\end{equation}
where\footnote{The incident wave derived by Ref. \cite{AB-1} is $e^{-iz \cos \phi - i \alpha \phi}$, which is different from that of Eq. (\ref{scattering-1}). The authors in this reference derived it by solving the 
appropriate differential equation. It was argued\cite{hagen-90-1} that this discrepancy is originated from the fact that the long-range nature of the vector potential does not allow the interchange of the 
summation over $m$ with the taking of the $r \rightarrow \infty$ limit in the partial-wave analysis.} $f_0 (\phi)$ is 
\begin{equation}
\label{scattering-2}
f_0 (\phi) = \frac{1}{\sqrt{2 \pi i k}} \left[ -2 \pi \delta (\phi - \pi) (1 - \cos \pi \alpha') - i e^{-i N (\phi - \pi)} \frac{\sin \pi \alpha' e^{-i \phi / 2}}{\cos \frac{\phi}{2}}   \right].
\end{equation}
In Eq. (\ref{scattering-2}) we used $\alpha' = N + \gamma$, where $N$ is integer and $0 \leq \gamma < 1$. 

Using 
\begin{equation}
\label{bessel-1}
\frac{1}{z} J_{\nu} (z) = \frac{1}{2 \nu} \left[ J_{\nu - 1} (z) + J_{\nu+ 1} (z)  \right],
\end{equation}
it is possible to show 
\begin{eqnarray}
\label{bessel-2}
&& F_2 (z; \mu, \nu) = \frac{1}{2 \nu} \left[ F_1 (z; \mu, \nu-1) + F_1 (z; \mu, \nu+1) \right]                      \\    \nonumber
&& F_3 (z; \mu, \nu) = \frac{1}{4 \mu \nu} \bigg[ F_1 (z; \mu - 1, \nu - 1) + F_1 (z; \mu - 1, \nu + 1)              \\    \nonumber
&& \hspace{4.0cm}  + F_1 (z; \mu + 1, \nu - 1) + F_1 (z; \mu + 1, \nu + 1) \bigg].
\end{eqnarray}
Then, $u_m (z)$ and $v_m (z)$ can be expressed as
\begin{eqnarray}
\label{wave-3}
&&u_m (z) = \frac{A_m \pi \hbar^2 k^2}{\sin (|m + \alpha'| \pi)} \Bigg[ - \frac{|m + \alpha'| \xi_m}{|m + \alpha'| + 1}  F_1 (z; -|m + \alpha'|, |m + \alpha'|)      \\   \nonumber                  
&& \hspace{6.0cm}+ \frac{\xi_m}{|m + \alpha'| + 1}  F_1 (z; -|m + \alpha'|, |m + \alpha'| + 2)                                                                                     \\   \nonumber
&& - \frac{\xi_{m,-}}{4 |m + \alpha'|^2} \bigg\{ F_1 (z; -|m + \alpha'|- 1, |m + \alpha'| - 1) + F_1 (z; -|m + \alpha'| + 1, |m + \alpha'| + 1)                                    \\    \nonumber
&& \hspace{1.0cm}  + F_1 (z; -|m + \alpha'|- 1, |m + \alpha'| + 1) + F_1 (z; -|m + \alpha'| + 1, |m + \alpha'| - 1)    \bigg\}   \Bigg]                                            \\    \nonumber
&&v_m (z) = -\frac{A_m \pi \hbar^2 k^2}{\sin (|m + \alpha'| \pi)} \Bigg[ - \frac{|m + \alpha'| \xi_m}{|m + \alpha'| + 1}  F_1 (z; |m + \alpha'|, |m + \alpha'|)      \\   \nonumber                  
&& \hspace{6.0cm}+ \frac{\xi_m}{|m + \alpha'| + 1}  F_1 (z; |m + \alpha'|, |m + \alpha'| + 2)                                                                                     \\   \nonumber
&& + \frac{\xi_{m,-}}{4 |m + \alpha'|^2} \bigg\{ F_1 (z; |m + \alpha'|- 1, |m + \alpha'| - 1) + F_1 (z; |m + \alpha'| + 1, |m + \alpha'| + 1)                                    \\    \nonumber
&& \hspace{4.0cm}  +  2F_1 (z; |m + \alpha'|- 1, |m + \alpha'| + 1)     \bigg\}   \Bigg].
\end{eqnarray}

Now, let us examine the behavior of $\psi ({\bm r})$ around $r \sim 0$. We use the following indefinite integral formula:
\begin{equation}
\label{iformula-1}
F_1 (z; \mu, \nu) = - \frac{z}{\mu^2 - \nu^2} \left[ J_{\mu+1} (z) J_{\nu} (z) - J_{\mu} (z) J_{\nu + 1} (z) \right] + \frac{1}{\mu + \nu} J_{\mu} (z) J_{\nu} (z).
\end{equation}
If we takes $\nu \rightarrow \pm \mu$ limit in Eq. (\ref{iformula-1}), one can also derive 
\begin{eqnarray}
\label{iformula-2}
&&F_1 (z; \mu, \mu) = \frac{2^{-1 - 2 \mu} z^{2 \mu}}{\mu \Gamma^2 (1 + \mu)} {_2F}_3 \left(\mu, \mu + \frac{1}{2}: 1 + \mu, 1 + \mu, 1 + 2 \mu: -z^2 \right)   \\   \nonumber
&&F_1 (z; \mu, -\mu) = - \frac{z^2} {4 \Gamma ( 2 - \mu) \Gamma (2 + \mu)}  {_3F}_4 \left(1, 1, \frac{3}{2}: 2 - \mu, 2 + \mu, 2, 2: -z^2 \right) + \frac{\ln z}{\Gamma (1 - \mu) \Gamma (1 + \mu)}
\end{eqnarray}
where $\Gamma (z)$ and $_pF_q (a_1, \cdots, a_p: b_1, \cdots, b_q: z)$ are the usual gamma and generalized hypergeometric functions. Using the limiting form 
\begin{equation}
\label{limiting-1}
\lim_{z \rightarrow 0} J_{\nu} (z) = \frac{1}{\Gamma (\nu + 1)} \left( \frac{z}{2} \right)^{\nu},
\end{equation}
one can show 
\begin{eqnarray}
\label{limiting-2}
&& \lim_{r \rightarrow 0} F_1 (z; \mu, \nu) = \frac{1}{(\mu + \nu) \Gamma (\mu + 1) \Gamma (\nu + 1)}  \left( \frac{z}{2} \right)^{\mu + \nu}      \\    \nonumber
&& \lim_{r \rightarrow 0} F_1 (z; \mu, \mu) = \frac{1}{2 \mu \Gamma^2 (1 + \mu)} \left( \frac{z}{2} \right)^{2 \mu}                                \\    \nonumber
&& \lim_{r \rightarrow 0} F_1 (z; \mu, -\mu) = \frac{\ln z}{\Gamma ( 1 - \mu) \Gamma (1 + \mu)}.
\end{eqnarray}
Then the dominant terms in $u_m$ and $v_m$ at $r \sim 0$ are
\begin{eqnarray}
\label{limiting-3}
&&\lim_{r \rightarrow 0} u_m (z) = - \frac{A_m \hbar^2 k^2}{8} \frac{\xi_{m, -}}{|m + \alpha'|} \left( \frac{z}{2} \right)^{-2}    \\   \nonumber
&&\lim_{r \rightarrow 0} v_m (z) = \frac{A_m \hbar^2 k^2}{8} \frac{\Gamma (-|m + \alpha'|) \xi_{m,-}}{(|m + \alpha'| - 1) \Gamma (1 + |m + \alpha'|)}  \left( \frac{z}{2} \right)^{2 (|m + \alpha'| - 1)},
\end{eqnarray}
which yield at ${\cal O} (\beta)$
\begin{eqnarray}
\label{limiting-4}
&&\lim_{r \rightarrow 0} \psi ({\bm r}) = \beta \sum_{m = -\infty}^{\infty} e^{i m \phi} \Bigg[ \frac{D_m}{\Gamma (1 - |m + \alpha'|)}  \left( \frac{z}{2} \right)^{- |m + \alpha'|}             \\      \nonumber
&& \hspace{4.0cm}  - \frac{A_m h^2 k^2}{8} \frac{\xi_{m,-}}{(|m + \alpha'| - 1) \Gamma (1 + |m + \alpha'|)} \left( \frac{z}{2} \right)^{|m + \alpha'| - 2} \Bigg].
\end{eqnarray}
Since we cannot make $\lim_{r \rightarrow 0} \psi ({\bm r})$ regular by choosing $D_m$ appropriately, unlike the AB-scattering in the usual quantum mechanics the AB-like scattering with GUP (\ref{GUP-d-4}) should allow the irregular solution at the origin.  

Now, let us examine the behavior of $\psi ({\bm r})$ around $r \sim \infty$. Using the limiting behavior of the Bessel function
\begin{equation}
\label{limiting-5}
\lim_{r \rightarrow \infty} J_{\nu} (z) = \sqrt{\frac{2}{\pi z}} \cos \left( z - \frac{\nu \pi}{2} - \frac{\pi}{4} \right) = \frac{1}{\sqrt{2 \pi z}} \left[ \left(-i \right)^{\nu + 1 / 2} e^{i z} + \left(i \right)^{\nu + 1 / 2} e^{-i z} \right],
\end{equation}
it is straightforward to show 
\begin{equation}
\label{limiting-6}
\lim_{r \rightarrow \infty} F_1(z; \mu, \nu) = - \frac{1}{2 \pi (\mu^2 - \nu^2)} \left[ \left\{ (-i)^{\mu - \nu +1} - (-i)^{\nu - \mu + 1} \right\} + \left\{ (-i)^{\nu - \mu - 1} - (-i)^{\mu - \nu - 1} \right\} \right].
\end{equation}
When $\nu = \pm \mu$ in Eq. (\ref{limiting-6}), one can derive the following asymptotic formula by making use of Eq. (\ref{iformula-2}):
\begin{equation}
\label{limiting-7}
\lim_{r \rightarrow \infty} F_1 (z; \mu, \mu) = \frac{1}{2 \mu}                               \hspace{.5cm}
\lim_{r \rightarrow \infty} F_1 (z; \mu, -\mu) = \frac{-\gamma - \psi \left( \frac{1}{2} \right) + \psi (1 - \mu) + \psi (1 + \mu)}{2 \Gamma (1 - \mu) \Gamma (1 + \mu)},
\end{equation}
where $\gamma$ and $\psi (z)$ are Euler number and digamma function. Using $\psi (z + 1) = \psi (z) + 1 / z$ explicitly, one can show 
\begin{equation}
\label{limiting-8}
\lim_{r \rightarrow \infty} u_m (z) = g_{1, m}   \hspace{2.0cm}  \lim_{r \rightarrow \infty} v_m (z) = g_{2, m},
\end{equation}
where
\begin{eqnarray}
\label{limiting-9}
&&  g_{1,m} = - \frac{A_m \hbar^2 k^2}{2 (1 + |m + \alpha'|)}  \Bigg[ \left( \xi_m + \frac{\xi_{m,-}}{2 |m + \alpha'| (1 - |m + \alpha'|)} \right) \bigg\{ - \gamma - \psi \left(\frac{1}{2} \right)    \\   \nonumber
&&\hspace{9.0cm} + \psi (|m + \alpha'|) + \psi (1 - |m + \alpha'|) \bigg\}                                                                                                                              \\   \nonumber
&& \hspace{2.5cm}  + \frac{1 + 2 |m + \alpha'|}{|m + \alpha'| (1 + |m + \alpha'|)} \xi_m - \frac{1 - |m + \alpha| - 3 |m + \alpha'|^2 + |m + \alpha|^3}{2 |m + \alpha'|^3 (1 + |m + \alpha'|) (1 - |m + \alpha'|)^2} \xi_{m, -}  \Bigg]    \\   \nonumber
&& g_{2, m} = \frac{A_m \hbar^2 k^2 \Gamma(|m + \alpha'|) \Gamma (1 - |m + \alpha'|)}{2 (1 + |m + \alpha'|)} \left( \xi_m + \frac{\xi_{m, -}}{2 |m + \alpha'| (1 - |m + \alpha'|)} \right).
\end{eqnarray}
Using Eq. (\ref{limiting-8}) it is straightforward to compute $\lim_{r \rightarrow \infty} f_{1,m} (r)$ explicitly. Since $f_{1.m} (r)$ should be outgoing wave at $r = \infty$, we should impose
the coefficient of $e^{-i k r}$ to be zero, which gives
\begin{equation}
\label{limiting-10}
C_m + g_{1,m} = - e^{-i \pi |m + \alpha'|} (D_m + g_{2, m}).
\end{equation}
Then,  $f_{1,m} (r \rightarrow \infty)$ reduces to 
\begin{equation}
\label{limiting-11}
\lim_{r \rightarrow \infty} f_{1,m} (r) = \frac{e^{i k r}}{\sqrt{2 \pi i k r}} e^{i  \pi|m + \alpha'| / 2} \left(1 - e^{-2 i \pi |m + \alpha'|} \right) (D_m + g_{2, m}).
\end{equation}
Thus, the asymptotic behavior of the wave function given in Eq. (\ref{wavef-1}) can be written as a standard from 
\begin{equation}
\label{limiting-12}
\lim_{r \rightarrow \infty} \psi ({\bm r}) = e^{-i k r \cos \phi} + \frac{e^{i k r}}{\sqrt{r}} f(\phi),
\end{equation}
where the scattering amplitude $f(\phi)$ is 
\begin{equation}
\label{s-amplitude-1}
f(\phi) = f_0 (\phi) + \beta f_1 (\phi) + {\cal O} (\beta^2).
\end{equation}
In Eq. (\ref{s-amplitude-1}) $f_0 (\phi)$ is given in Eq. (\ref{scattering-2}) and $f_1 (\phi)$ is 
\begin{equation}
\label{s-amplitude-2}
f_1 (\phi) = \frac{1}{\sqrt{2 \pi i k}} \sum_{m = -\infty}^{\infty} e^{i m \phi}  e^{i |m + \alpha'| \pi / 2} \left(1 - e^{-2 i \pi |m + \alpha'|} \right) (D_m + g_{2, m}).
\end{equation}

Here, we consider a special case $D_m = 0$ for all $m$. Inserting $\xi_m$ and $\xi_{m, -}$ given in Eq. (\ref{radial-7}) into $g_{2, m}$, one can express $g_{2, m}$ in a form;
\begin{eqnarray}
\label{special-1}
&&g_{2, m} = - \frac{(-i)^{|m + \alpha'|} \hbar^2 k^2}{4} \Gamma (m + \alpha') \Gamma (-m - \alpha')                   \\    \nonumber
&& \hspace{2.0cm}   \times \left[ -2 \alpha'^2 + 6 \alpha' (m + \alpha') + \alpha'^2 (m + \alpha') \left\{ \frac{1 - \alpha' / 2}{1 - m - \alpha'} - \frac{1 + \alpha' / 2}{1 + m + \alpha'} \right\} \right].
\end{eqnarray}
Here, we used $A_m = (-i)^{|m + \alpha'|}$. It is worthwhile noting that except $A_m$ there is no absolute value in Eq. (\ref{special-1}). 
Inserting Eq. (\ref{special-1}) into Eq. (\ref{s-amplitude-2}), we get
\begin{eqnarray}
\label{special-2}
&& f_1 (\phi) = - \frac{\hbar^2 k^2}{4} \frac{1}{\sqrt{2 \pi i k}}  \sum_{m = -\infty}^{\infty} e^{i m \phi}  \left(1 - e^{-2 i \pi |m + \alpha'|} \right)                  \\    \nonumber
&& \hspace{.3cm}   \times \Bigg[ -2 \alpha'^2 \Gamma (m + \alpha') \Gamma (-m - \alpha')  + 6 \alpha' \Gamma (1 + m + \alpha') \Gamma (-m - \alpha')                       \\    \nonumber
&& \hspace{1.0cm}+ 
\alpha'^2  \left( 1 - \frac{\alpha'}{2} \right) \frac{\Gamma (1 + m + \alpha') \Gamma (-m - \alpha')}{1 - m - \alpha} - \alpha'^2  \left( 1 + \frac{\alpha'}{2} \right) \frac{\Gamma (1 + m + \alpha') \Gamma (-m - \alpha')}{1 + m + \alpha} \Bigg].
\end{eqnarray}
Let us express $\alpha'$ as $\alpha' = N + \gamma$, where $N$ is integer and $0 \leq \gamma < 1$. Then, $f_1 (\phi)$ in Eq. (\ref{special-2}) can be written as following form:
\begin{eqnarray}
\label{special-3}
&& f_1 (\phi)                                                                      \\    \nonumber
&&= - \frac{i \hbar^2 k^2}{2} \sin (\pi \gamma) \frac{e^{-i \pi \gamma}}{\sqrt{2 \pi i k}} \left[ -2 \alpha'^2 J_1 + 6 \alpha' J_2 + \alpha'^2 \left( 1 - \frac{\alpha'}{2} \right) J_3 - \alpha'^2 \left( 1 + \frac{\alpha'}{2} \right) J_4 \right]   \\   \nonumber
&&+  \frac{i \hbar^2 k^2}{2} \sin (\pi \gamma) \frac{e^{i \pi \gamma}}{\sqrt{2 \pi i k}} \left[ -2 \alpha'^2 K_1 + 6 \alpha' K_2 + \alpha'^2 \left( 1 - \frac{\alpha'}{2} \right) K_3 - \alpha'^2 \left( 1 + \frac{\alpha'}{2} \right) K_4 \right],
\end{eqnarray}
where
\begin{eqnarray}
\label{special-4}
&&J_1 = \sum_{m = -N}^{\infty} e^{i m \phi} \Gamma (m + \alpha') \Gamma (-m - \alpha')       \hspace{.3cm}
J_2 = \sum_{m = -N}^{\infty} e^{i m \phi} \Gamma (1 + m + \alpha') \Gamma (-m - \alpha')                               \\    \nonumber
&&J_3 = \sum_{m = -N}^{\infty} e^{i m \phi} \frac{\Gamma (1 + m + \alpha') \Gamma(-m - \alpha')}{1 - m - \alpha'}      \hspace{.3cm}
J_4 = \sum_{m = -N}^{\infty} e^{i m \phi} \frac{\Gamma (1 + m + \alpha') \Gamma(-m - \alpha')}{1 + m + \alpha'}
\end{eqnarray}
and $K_j \hspace{.1cm} (j = 1, 2, 3, 4)$ are similar to $J_j$. The only difference is the summation range, which is from $-\infty$ to $-N - 1$. 
It is straightforward to show 
\begin{eqnarray}
\label{special-5}
&&K_1 = J_1 \Bigg|_{\phi \rightarrow - \phi, \alpha' \rightarrow -\alpha'}   \hspace{1.0cm} K_2 = -J_2 \Bigg|_{\phi \rightarrow - \phi, \alpha' \rightarrow -\alpha'}     \\    \nonumber
&& K_3 = -J_4 \Bigg|_{\phi \rightarrow - \phi, \alpha' \rightarrow -\alpha'} \hspace{1.0cm}  K_4 = -J_3 \Bigg|_{\phi \rightarrow - \phi, \alpha' \rightarrow -\alpha'}.
\end{eqnarray} 
Of course, $\alpha' \rightarrow - \alpha'$ implies $N \rightarrow -N - 1$ and $\gamma \rightarrow 1 - \gamma$. Then, it is easy to show that $f_1 (\phi)$ is invariant under the 
simultaneous change $\phi \rightarrow -\phi$, $\alpha' \rightarrow - \alpha'$, which is a symmetry of the Hamiltonian.  Summing over $m$, one can show
\begin{eqnarray}
\label{special-6}
&&J_1 = - \frac{\pi}{\gamma \sin (\pi \gamma)} e^{- i N \phi} {_2 F}_1 (1, \gamma: 1 + \gamma: -e^{i \phi})           \\    \nonumber
&& \hspace{.6cm} = - \frac{\pi e^{-i (N + 1/2) \phi}}{2 \gamma \sin (\pi \gamma) \cos(\phi / 2)} {_2 F}_1 \left(1, 1: 1 + \gamma: \frac{e^{i \phi / 2}}{2 \cos (\phi/2)} \right)      \\    \nonumber
&&J_2 = - \frac{\pi}{\sin (\pi \gamma)} \frac{e^{-i N \phi}}{1 + e^{i \phi}} = - \frac{\pi e^{-i (N + 1 / 2) \phi}}{2 \sin (\pi \gamma) \cos (\phi / 2)}                              \\    \nonumber
&&J_3 = - \frac{\pi}{(1 - \gamma) \sin (\pi \gamma)} e^{-i N \phi} {_2 F}_1 (1, -1 + \gamma: \gamma: -e^{i \phi})                                                                     \\    \nonumber
&& \hspace{.6cm} = - \frac{\pi e^{-i (N + 1/2) \phi}}{2 (1 - \gamma) \sin (\pi \gamma) \cos(\phi / 2)} {_2 F}_1 \left(1, 1: \gamma: \frac{e^{i \phi / 2}}{2 \cos (\phi/2)} \right)      \\    \nonumber
&&J_4 = - \frac{\pi}{(1 + \gamma) \sin (\pi \gamma)} e^{-i N \phi} {_2 F}_1 (1, 1 + \gamma: 2 + \gamma: -e^{i \phi})                                                                     \\    \nonumber
&& \hspace{.6cm} = - \frac{\pi e^{-i (N + 1/2) \phi}}{2 (1 + \gamma) \sin (\pi \gamma) \cos(\phi / 2)} {_2 F}_1 \left(1, 1: 2 + \gamma: \frac{e^{i \phi / 2}}{2 \cos (\phi/2)} \right)
\end{eqnarray}
where $_2 F_1 (a, b: c: z)$ is a hypergeometric function and we used the identity
\begin{equation}
\label{special-7}
_2 F_1 (a, b: c: z) = (1 - z)^{-a} {_2 F}_1 \left( a, c - b: c: \frac{z}{z - 1} \right).
\end{equation}
Using Eq. (\ref{special-5}) and Eq. (\ref{special-6}) it is straightforward to show 
\begin{eqnarray}
\label{special-8}
&&K_1 = - \frac{\pi e^{-i (N + 1/2) \phi}}{2 (1 - \gamma) \sin (\pi \gamma) \cos(\phi / 2)} {_2 F}_1 \left(1, 1: 2 - \gamma: \frac{e^{-i \phi / 2}}{2 \cos (\phi/2)} \right)             \\    \nonumber
&&K_2 =  \frac{\pi e^{-i (N + 1 / 2) \phi}}{2 \sin (\pi \gamma) \cos (\phi / 2)}                                                                                                         \\    \nonumber
&&K_3 =  \frac{\pi e^{-i (N + 1/2) \phi}}{2 (2 - \gamma) \sin (\pi \gamma) \cos(\phi / 2)} {_2 F}_1 \left(1, 1: 3 - \gamma: \frac{e^{-i \phi / 2}}{2 \cos (\phi/2)} \right)              \\    \nonumber
&&K_4 =  \frac{\pi e^{-i (N + 1/2) \phi}}{2 \gamma \sin (\pi \gamma) \cos(\phi / 2)} {_2 F}_1 \left(1, 1: 1 - \gamma: \frac{e^{-i \phi / 2}}{2 \cos (\phi/2)} \right). 
\end{eqnarray}
Inserting Eqs (\ref{special-6}) and (\ref{special-8}) into Eq. (\ref{special-3}), one can show 
\begin{equation}
\label{special-9}
f_1 (\phi) = \frac{i \pi \hbar^2 k^2 e^{-i (N + 1 / 2) \phi}}{4 \cos (\phi / 2) \sqrt{2 \pi i k}} G(\alpha', \phi)
\end{equation}
where 
\begin{eqnarray}
\label{special-10}
&&G(\alpha', \phi)                                                                                                \\    \nonumber
&&= 2 \alpha'^2 \left[ \frac{e^{i \pi \gamma}}{1 - \gamma} {_2 F}_1 (1, 1: 2 - \gamma: x^*) - \frac{e^{-i \pi \gamma}}{\gamma} {_2 F}_1 (1, 1: 1 + \gamma: x)  \right] + 12 \alpha' \cos (\pi \gamma)   \\   \nonumber
&& \hspace{1.0cm} + \alpha'^2 \left(1 - \frac{\alpha'}{2} \right) \left[ \frac{e^{i \pi \gamma}}{2 - \gamma} {_2 F}_1 (1, 1: 3 - \gamma: x^*) + \frac{e^{-i \pi \gamma}}{1 - \gamma} {_2 F}_1 (1, 1: \gamma: x) \right]   \\   \nonumber
&& \hspace{1.0cm} -  \alpha'^2 \left(1 + \frac{\alpha'}{2} \right) \left[ \frac{e^{i \pi \gamma}}{\gamma} {_2 F}_1 (1, 1: 1 - \gamma: x^*) + \frac{e^{-i \pi \gamma}}{1 + \gamma} {_2 F}_1 (1, 1: 2 + \gamma: x) \right]. 
\end{eqnarray}
In Eq. (\ref{special-10}) $x$ is given by 
\begin{equation}
\label{special-11}
x = \frac{e^{i \phi / 2}}{2 \cos (\phi / 2)}
\end{equation}
and $x^*$ is its complex conjugate. From Eq. (\ref{special-9}) one can show again that $f_1 (\phi)$ is invariant  under the 
simultaneous change $\phi \rightarrow -\phi$, $\alpha' \rightarrow - \alpha'$.
If $\phi \neq \pi$, the scattering amplitude becomes
\begin{equation}
\label{special-12}
f (\phi) = \frac{-i e^{-i (N + 1/2) \phi}}{\cos (\phi / 2) \sqrt{2 \pi i k}} \left[ \sin (\pi \gamma) - \frac{\pi \hbar^2 k^2 \beta}{4} G(\alpha', \phi) + {\cal O} (\beta^2) \right].
\end{equation}
Then, the differential cross section reduces to 
\begin{eqnarray}
\label{special-13}
&& \frac{d \sigma}{d \phi} = \frac{1}{2 \pi k \cos^2 (\phi / 2)} \Bigg|\sin (\pi \gamma) - \frac{\pi \hbar^2 k^2 \beta}{4} G(\alpha', \phi) \Bigg|^2 + {\cal O} (\beta^2)    \\   \nonumber
&& \hspace{.7cm} = \frac{\sin (\pi \gamma)}{2 \pi k \cos^2 (\phi / 2)} \left[ \sin (\pi \gamma) - \beta \frac{\pi \hbar^2 k^2}{2} \mbox{Re} \hspace{.1cm} G(\alpha', \phi) \right] + {\cal O} (\beta^2).
\end{eqnarray}

\begin{figure}[ht!]
\begin{center}
\includegraphics[height=5.0cm]{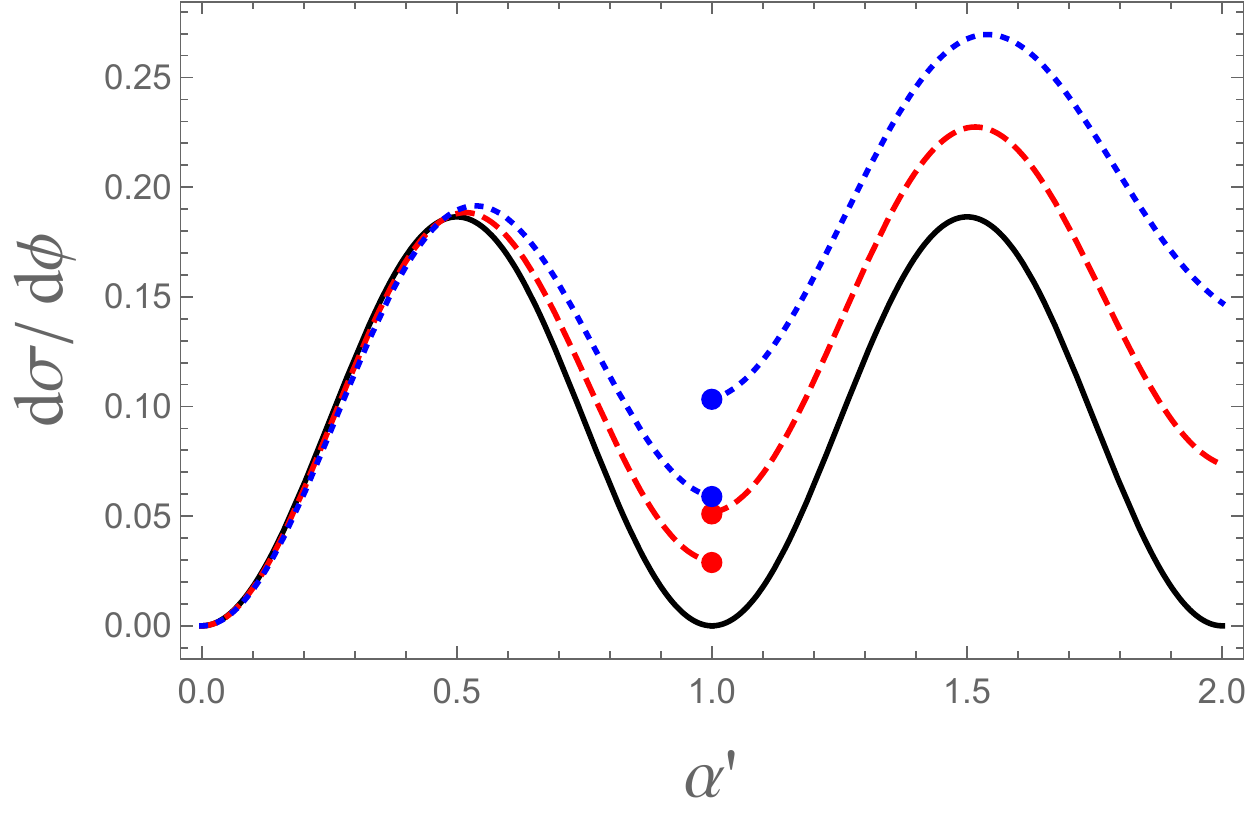} 
\includegraphics[height=5.0cm]{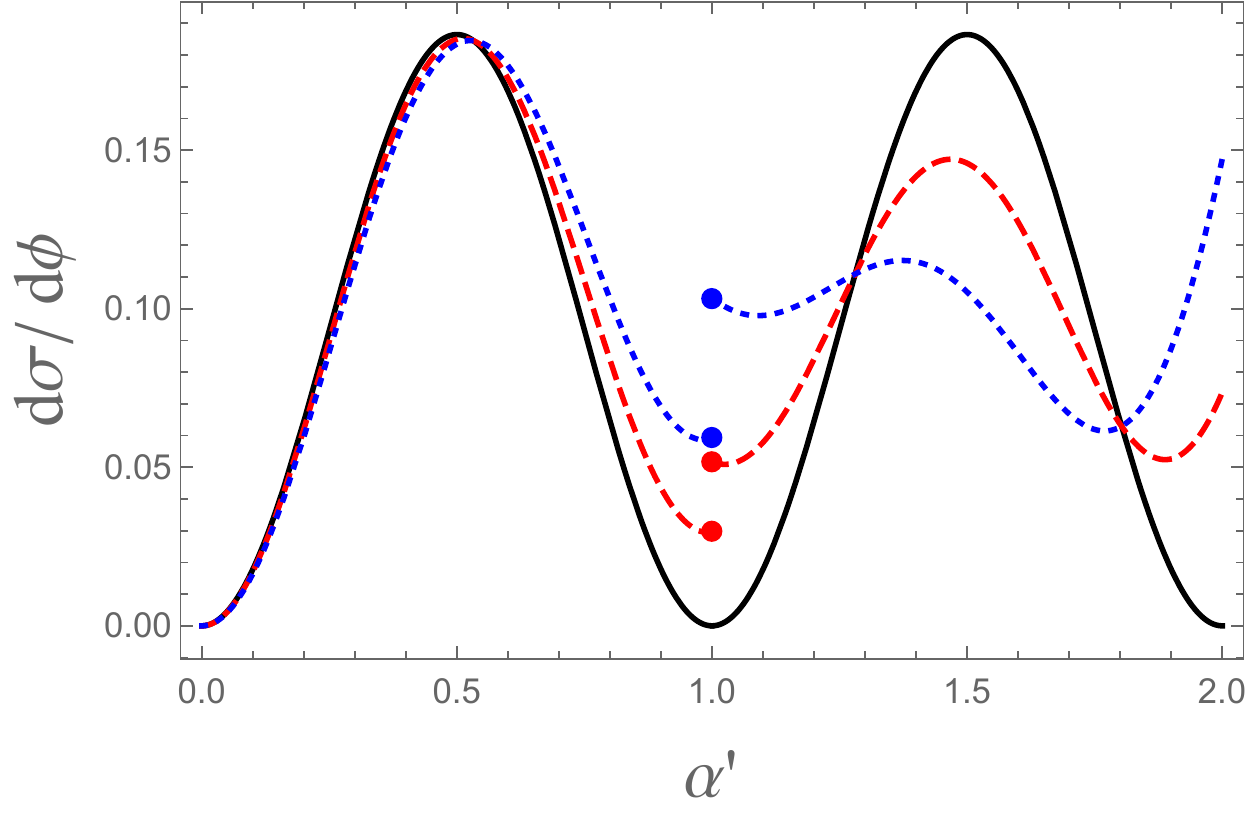}
\caption[fig2]{(Color online) The $\alpha'$-dependence of the differential cross section when $\beta = 0$ (black solid line), $\beta = 0.01$ (red dashed line), and $\beta =0.02$ (blue dotted line) for $\phi = \pi / 4$ (Fig. 2(a)) and $\phi = - \pi / 4$ (Fig. 2(b)). 
We chose $\hbar = k = 1$ for simplicity. As expected Fig. 2 exhibits discontinuous behavior at $\alpha' = 1$ when $\beta \neq 0$.
 }
\end{center}
\end{figure}

In the usual quantum mechanics with HUP the differential cross section vanishes when $\alpha'$ is integer. This is analogous to the Ramsauer effect\cite{bohm1951}. 
However, this behavior is not maintained at the first order of $\beta$. Furthermore, discontinuity occurs at every integer of $\alpha'$. For example, if 
$\alpha' = N^+ = \lim_{\gamma \rightarrow 0} (N + \gamma)$, one can show from the second expression of Eq. (\ref{special-13})
\begin{equation}
\label{discontinuity-1}
\frac{d \sigma}{d \phi} = \beta \frac{\pi \hbar^2 k N^2}{2} \left[ 2 (N - 2) \cos^2 \frac{\phi}{2} + 6 - N \right].
\end{equation}
If, however, $\alpha' = N^- = \lim_{\gamma \rightarrow 1} [(N - 1) + \gamma]$, one can also show
\begin{equation}
\label{discontinuity-2}
\frac{d \sigma}{d \phi} = \beta \frac{\pi \hbar^2 k N^2}{2} \left[ N + 6 - 2 (N + 2) \cos^2 \frac{\phi}{2} \right].
\end{equation}
In Fig. 2 we plot the $\alpha'$-dependence of the differential cross section when $\beta = 0$ (black solid line), $\beta = 0.01$ (red dashed line), and $\beta =0.02$ (blue dotted line) for $\phi = \pi / 4$ (Fig. 2(a)) and $\phi = - \pi / 4$ (Fig. 2(b)). 
We chose $\hbar = k = 1$ for simplicity.
As expected Fig. 2 exhibits discontinuous behavior at $\alpha' = 1$ when $\beta \neq 0$. Another interesting behavior Fig. 2 shows is the fact that while $d \sigma / d \phi$ at $\phi = \pm \theta$ are exactly identical in the usual 
quantum mechanics, this symmetry is obviously broken due to $G(\alpha', \phi)$.

\begin{figure}[ht!]
\begin{center}
\includegraphics[height=5.0cm]{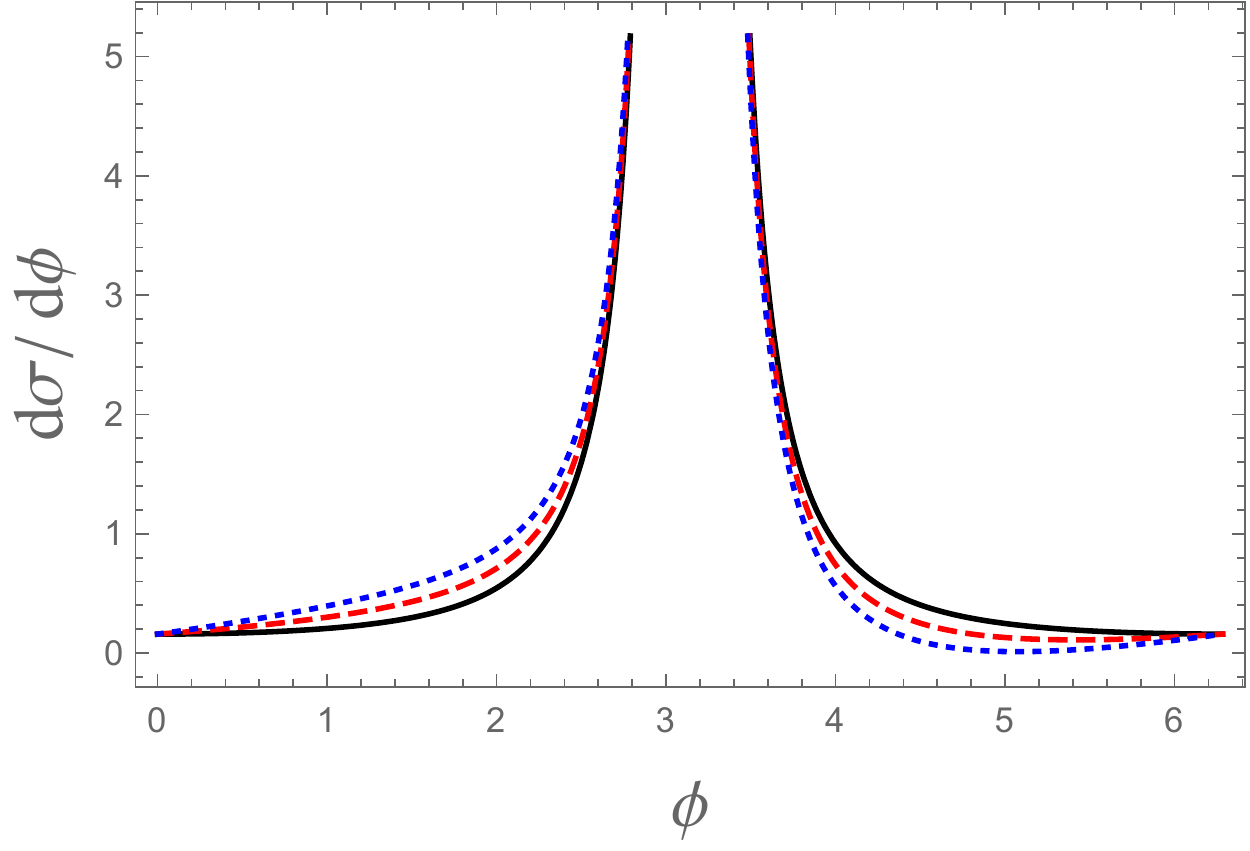} 
\caption[fig3]{(Color online) The $\phi$-dependence of the differential cross section when $\beta = 0$ (black solid line), $\beta = 0.004$ (red dashed line), and $\beta =0.008$ (blue dotted line). 
We chose $\hbar = k = 1$ and $\alpha' = 2.5$ for simplicity. Fig. 3 shows apparently that the symmetry (\ref{symm-10}) is broken when $\beta \neq 0$.
 }
\end{center}
\end{figure}

In usual quantum mechanics with HUP the cross section is symmetric at $\phi = \pi$, i.e.
\begin{equation}
\label{symm-10}
\frac{d \sigma}{d \phi} (\phi = \pi - \theta) = \frac{d \sigma}{d \phi} (\phi = \pi + \theta).
\end{equation}
However, this symmetry is also broken at the first order of $\beta$ because 
\begin{equation}
\label{symm-11}
x (\phi = \pi - \theta) = x^* (\phi = \pi + \theta) = \frac{i e^{-i \theta / 2}}{2 \sin (\theta / 2)}
\end{equation}
and $\mbox{Re}  \hspace{.1cm} G(\alpha', \phi) $ does not have $x \leftrightarrow x^*$ symmetry.
In order to confirm the fact we plot the $\phi$-dependence of the differential cross section when $\beta = 0$ (black solid line), $\beta = 0.004$ (red dashed line), and $\beta =0.008$ (blue dotted line) in Fig. 3. 
We chose $\hbar = k = 1$ and $\alpha' = 2.5$ for simplicity. This figure obviously show that the symmetry (\ref{symm-10}) is broken when $\beta \neq 0$. 

\section{Conclusion}

\begin{center}
\begin{tabular}{c|c|c} \hline \hline
  &  HUP  &  GUP                                                               \\   \hline  
$\phi \rightarrow -\phi, \alpha' \rightarrow - \alpha'$ symmetry & Y    & Y                             \\ 
symmetry of $\frac{d \sigma}{d \phi}$ at $\phi = \pi$ &  Y   &   N                                      \\    
Ramsauer effect & Y    &    N and discontinuous at integer $\alpha'$             \\    \hline
\end{tabular}

\vspace{0.3cm}
Table I: Comparison between usual and GUP-corrected AB-like effect
\end{center}
\vspace{0.5cm}

In this paper we explored how the Aharonov-Bohm scattering is modified in the GUP-corrected quantum mechanics. 
In Table I we compare the GUP-correct AB-like phenomenon with the usual AB-effect. 
The most striking difference is that the cross section is discontinuous at every integer $\alpha'$ due to $G(\alpha', \phi)$ given in Eq. (\ref{special-10}). From Eqs. (\ref{discontinuity-1}) and (\ref{discontinuity-2}) one can show that 
the discontinuity width at $\alpha' = N$ is 
\begin{equation}
\label{width}
\Delta \frac{d \sigma}{d \phi} = \beta \pi \hbar^2 k N^3 \big| 2 \cos^2 \frac{\phi}{2} - 1 \big|.
\end{equation}
Thus, it is possible, in principle, to verify the presence or absence of GUP experimentally by measuring the discontinuity. 
Of course, it seems to be very difficult to measure it because the discontinuity arises at the order of $\beta$, and $\beta$ is 
believed to be extremely small. 

One can use the Lagrangian (\ref{classical-4}) to derive the Feynman propagator (or Kernel) for the spin-$0$ AB-system with GUP. The propagator of the usual AB-system was 
derived long ago in Ref. \cite{inomata,gerry,yoo}. It seems to be of interest to explore the quantum effect by deriving the Feynman propagator corresponding to the Lagrangian (\ref{classical-4}). 
In the usual quantum mechanics it is well-known that 
the magnetic flux trapped by a superconductor ring is quantized by $\Phi = \pi n \hbar c / e \hspace{.2cm} (n = 0, \pm 1, \pm 2, \cdots)$. It is of interest to explore how this quantization rule is modified in the 
presence of GUP. 

Finally, one can extend this paper to the spin-$1/2$ AB problem in the presence of GUP. As commented earlier the Zeeman interaction term in this case is expressed as a $2$-dimensional singular $\delta$-function potential. 
One dimensional $\delta$-function potential problem in the GUP-corrected quantum mechanics was recently discussed in Ref. \cite{park2020}. It was shown in this reference that unlike usual quantum mechanics, the Schr\"{o}dinger 
and Feynman's path-integral approaches are inequivalent at the first order of $\beta$. It seems to be of interest to examine whether the $2$-dimensional $\delta$-function potential yields a similar result or not 
in the spin-$1/2$ AB problem with GUP.

\vspace{1.0cm}

{\bf Acknowledgments}:

This work was supported by the Kyungnam University Foundation Grant, 2020.


\begin{thebibliography}{99}
\bibitem{mead64} C. A. Mead, {\it Possible Connection Between Gravitation and Fundamental Length}, Phys. Rev. {\bf 135} (1964) B849.
\bibitem{townsend76} P. K. Townsend, {\it Small-scale structure of spacetime as the origin of the gravitational constant}, Phys. Rev. {\bf D 15} (1977) 2795.
\bibitem{amati89} D. Amati, M. Ciafaloni, and G. Veneziano, {\it Can spacetime be probed below the string size?}, Phys. Lett. {\bf B 216} (1989) 41.
\bibitem{garay94} L. J. Garay, {\it Quantum gravity and minimum length}, Int. J. Mod. Phys. {\bf A 10} (1995) 145 [gr-qc/9403008].
\bibitem{rovelli98} C. Rovelli, {\it Loop Quantum Gravity}, Living Rev. Relativity, {\bf 1} (1998) 1 [gr-qc/9710008].
\bibitem{carlip01} S. Carlip, {\it Quantum Gravity: a Progress Report}, Rep. Prog. Phys. {\bf 64} (2001) 885 [gr-qc/0108040].
\bibitem{konishi90} K. Konishi, G. Paffuti, and P. Provero, {it Minimum physical length and the generalized uncertainty principle in string theory}, Phys. Lett. {\bf B 234} (1990) 276.
\bibitem{kato90} M. Kato, {\it Particle theories with minimum observable length and open string theory}, Phys. Lett. {\bf B 245} (1990) 43.
\bibitem{padmanabhan85} T. Padmanabhan, {\it Physical significance of planck length}, Ann. Phys. {\bf 165} (1985) 38.
\bibitem{padmanabhan87} T. Padmanabhan, {\it Limitations on the operational definition of spacetime events and quantum gravity}, Class. Quant. Grav. {\bf 4} (1987) L107.
\bibitem{greensite91} J. Greensite, {Is there a minimum length in D=4 lattice quantum gravity?}, Phys. Lett. {\bf B 255} (1991) 375.
\bibitem{maggiore93}  M. Maggiore, {\it A Generalized Uncertainty Principle in Quantum Gravity}, Phys. Lett. {\bf B 304} (1993) 65 [hep-th/9301067].
\bibitem{uncertainty} W. Heisenberg, \"{U}ber den anschaulichen Inhalt der quantentheoretischen Kinematik und Mechanik, Z. Phys. {\bf 43} (1927) 172.
\bibitem{robertson1929} H. P. Robertson, {\it The Uncertainty Principle}, Phys. Rev. {\bf 34} (1929) 163. 
\bibitem{kempf93} A. Kempf, {\it Uncertainty Relation in Quantum Mechanics with Quantum Group Symmetry}, J. Math. Phys. {\bf 35} (1994) 4483 [hep-th/9311147].
\bibitem{kempf94}  A. Kempf, G. Mangano, and R. B. Mann, {\it Hilbert Space Representation of the Minimal Length Uncertainty Relation}, Phys. Rev. {\bf D 52} (1995) 1108 [hep-th/9412167].

\bibitem{scar99-1} F. Scardigli, {\it Generalized uncertainty principle in quantum gravity from micro-black hole gedanken experiment}, Phys. Lett, {\bf B 452} (1999) 39 [hep-th/9904025].
\bibitem{adler99-1} R. J. Adler and D. I. Santiago, {\it On Gravity and The Uncertainty Principle}, Mod. Phys. Lett. {\bf A 14} (1999) 1371 [gr-qc/9904026].
\bibitem{okamura02-1} L. N. Chang, D. Minic, N. Okamura, and T. Takeuchi, {\it Effect of the minimal length uncertainty relation on the density of states and the cosmological constant problem}, Phys. Rev. {\bf D 65} (2002) 125028 [hep-th/0201017].
\bibitem{okamura02-2} S. Benczik, L. N. Chang, D. Minic, N. Okamura, S. Rayyan, and T. Takeuchi, {\it Short distance versus long distance physics: The classical limit of the minimal length uncertainty relation}, Phys. Rev. {\bf D 66} (2002) 026003 [hep-th/0204049].
\bibitem{scar10-1} P. Jizba, H. Kleinert, and F. Scardigli, {\it Uncertainty relation on a world crystal and its applications to micro black holes}, Phys. Rev. {\bf D 81} (2010) 084030 [arXiv: 0912.2253 (hep-th)].
\bibitem{scar12-1} P. Jizba and F. Scardigli, {\it Emergence of special and doubly special relativity}, Phys. Rev. {\bf D 86} (2012) 025029 [arXiv:1105.3930 (hep-th)].
\bibitem{scar15-1} F. Scardigli and R. Casadio, {\it Gravitational tests of the generalized uncertainty principle}, Eur. Phys. J. {\bf C 75} (2015) 425 [arXiv:1407.0113 (hep-th)].







\bibitem{feynman} R. P. Feynman and A. R. Hibbs, Quantum Mechanics and Path Integrals (McGraw-Hill, 1965, New York).
\bibitem{kleinert} H. Kleinert, Path integrals in Quantum Mechanics, Statistics, and Polymer Physics (World Scientific,1995,  Singapore).
\bibitem{das2012} S. Das and S. Pramanik, {\it Path Integral for non-relativistic Generalized Uncertainty Principle corrected Hamiltonian}, Phys. Rev. {\bf D 86} (2012) 085004 [arXiv:1205.3919 (hep-th)].
\bibitem{gangop2019}S. Gangopadhyay and S. Bhattacharyya, {\it Path-integral action of a particle with the generalized uncertainty principle and correspondence with noncommutativity}, Phys. Rev. {\bf D 99} (2019) 104010 [arXiv:1901.03411 (quant-ph)].
\bibitem{comment-1} DaeKil Park and Eylee Jung, {\it Comment on “Path-integral action of a particle with the generalized uncertainty principle and correspondence with noncommutativity''}, Phys. Rev. {\bf D 101} (2020) 068501 [arXiv:2002.07954 (quant-ph)].
\bibitem{park20-1} DaeKil Park, ``Generalized uncertainty principle and $d$-dimensional quantum mechanics'', Phys. Rev. 
{\bf D 101} (2020) 106013 [arXiv:2003.13856 (quant-ph)].






















\bibitem{AB-1} Y. Aharonov and D. Bohm, {\it Significance of Electromagnetic Potentials in the Quantum Theory}, Phys. Rev. {\bf 115} (1959) 485.
\bibitem{hagen-91} C. R. Hagen, {\it Spin Dependence of the Aharonov-Bohm Effect}, Int. J. Mod. Phys. {\bf A 6} (1991) 3119.
\bibitem{peshkin} M. Peshkin and A. Tonomura, {\it The Aharonov-Bohm Effect} (Springer-Verlag, Berlin, 1989).
\bibitem{hagen-90-2} C. R. Hagen, {\it Aharonov-Bohm Scattering of Particles with Spin}, Phys. Rev. Lett. {\bf 64} (1990) 503. 
\bibitem{park-95} D. K, Park, {\it Green's function approach to two- and three-dimensional delta-function potentials and application to the spin-1/2 Aharonov-Bohm problem}, J. Math. Phys. {\bf 36} (1995) 5453.
\bibitem{jackiw} R. Jackiw, {\it Delta-funstion potential in two- and three-dimensional quantum mechanics}, in M. A. B\'eg memorial volume, A. Ali and P. Hoodbhoy, eds. (World Scientific, Singapore, 1991).
\bibitem{capri} A. Z. Capri, {\it Nonrelativistic Quantum Mechanics} (Benjamin/Cummings, 1985,  Menlo Park).
\bibitem{huang} H. Huang, {\it Quarks, Leptons, and Gauge Fields} (World Scientific, 1982, Singapore).
\bibitem{park97-1} D. K. Park and S. K. Yoo, {\it Equivalence of renormalization with self-adjoint extension in Green's function formalism}, hep-th/9712134 (unpublished).
\bibitem{hagen-90-1} C. R. Hagen, {\it Aharonov-Bohm scattering amplitude}, Phys. Rev. {\bf D 41} (1990) 2015.
\bibitem{bohm1951} D. Bohm, {\it Quantum Theory} (Prentice-Hall, Englewood Cliffs, New Jersey, 1951).

\bibitem{inomata} D. Peak and A. Inomata, {\it Summation over Feynman Histories in Polar Coordinates}, J. Math. Phys. {\bf 10} (1969) 1422.
\bibitem{gerry} C. C. Gerry and V. A. Singh, {\it Feynman path-integral approach to the Aharonov-Bohm effect}, Phys. Rev. {\bf D 20} (1979) 2550.
\bibitem{yoo} D. K. Park and S. K. Yoo, {\it Propagator for spinless and spin-1/2 Aharonov-Bohm-Coulomb systems}, Ann. Phys. {\bf 263} (1998) 295 [hep-th/9707024].
\bibitem{park2020}  D. K. Park and Eylee Jung, ``Generalized Uncertainty Principle and Point Interaction'', Phys. Rev. {\bf D 101} (2020) 066007 [arXiv:2001.02850 (quant-ph)].  







\end{thebibliography}
\end{document}